\documentclass[%
 reprint,
superscriptaddress,
 amsmath,amssymb,
pra,
]{revtex4-1}
\usepackage{longtable}
\usepackage{ulem}
\usepackage{multirow}
\usepackage{graphicx}
\usepackage{dcolumn}
\usepackage{bm}
\usepackage{booktabs}
\usepackage{enumerate}
\usepackage{color}
\newcommand{\te}[1]{\textrm{#1}}

\newcommand{\balpha}{\boldsymbol{\alpha}}

\begin{document}

\preprint{APS/123-QED}

\title{Benchmarking calculations with spectroscopic accuracy of excitation energies and wavelengths in sulfur-like tungsten}

\author{Chun Yu Zhang}
\affiliation{Shanghai EBIT Lab, Key Laboratory of Nuclear Physics and Ion-beam Application, Institute of Modern Physics, Department of Nuclear Science and Technology, Fudan University, Shanghai 200433, Peoples Republic of China;}%
\affiliation{Spectroscopy, Quantum Chemistry and Atmospheric Remote Sensing (SQUARES), CP160/09, Universit\'{e} libre de Bruxelles, Av. F.D. Roosevelt 50, 1050 Brussels, Belgium;}

\author{Kai Wang}
\email{rsi13@fudan.edu.cn; wang$_$kai10@fudan.edu.cn}
\affiliation{Spectroscopy, Quantum Chemistry and Atmospheric Remote Sensing (SQUARES), CP160/09, Universit\'{e} libre de Bruxelles, Av. F.D. Roosevelt 50, 1050 Brussels, Belgium;}
\affiliation{Hebei Key Lab of Optic-electronic Information and Materials, The College of Physics Science and Technology, Hebei University, Baoding 071002, Peoples Republic of China;}

\author{Michel Godefroid}
\affiliation{Spectroscopy, Quantum Chemistry and Atmospheric Remote Sensing (SQUARES), CP160/09, Universit\'{e} libre de Bruxelles, Av. F.D. Roosevelt 50, 1050 Brussels, Belgium;}

\author{Per J{\"o}nsson}
\affiliation{Department of Materials Science and Applied Mathematics, Malm{\"o} University, SE-20506, Malm{\"o}, Sweden;}
\author{Ran Si}
\email{rsi13@fudan.edu.cn; wang$_$kai10@fudan.edu.cn}
\affiliation{Shanghai EBIT Lab, Key Laboratory of Nuclear Physics and Ion-beam Application, Institute of Modern Physics, Department of Nuclear Science and Technology, Fudan University, Shanghai 200433, Peoples Republic of China;}%
\affiliation{Spectroscopy, Quantum Chemistry and Atmospheric Remote Sensing (SQUARES), CP160/09, Universit\'{e} libre de Bruxelles, Av. F.D. Roosevelt 50, 1050 Brussels, Belgium;}

\author{Chong Yang Chen}%
\affiliation{Shanghai EBIT Lab, Key Laboratory of Nuclear Physics and Ion-beam Application, Institute of Modern Physics, Department of Nuclear Science and Technology, Fudan University, Shanghai 200433, Peoples Republic of China;}%

\date{\today}

\begin{abstract}
Atomic properties of S-like W are evaluated through a state-of-the-art method, namely, the multi-configuration Dirac-Hartree-Fock (MCDHF) method combined with the relativistic configuration interaction (RCI) approach. The level energies, wavelengths, and transition parameters involving the 88 lowest levels of W$^{+58}$ (W LIX) are calculated.
We discuss in detail the relative importance of the valence- and core-valence electron correlation effects, the Breit interaction, the higher order retardation correction beyond the Breit interaction through the transverse photon interaction, and the quantum electrodynamical (QED) corrections.
The present level energies are highly accurate, with uncertainties close to what can be achieved  from spectroscopy.	
As such, they provide benchmark tests for other theoretical calculations of  S-like W  and should assist the spectroscopists in their assignment/identification  of observed lines in complex spectra.
\end{abstract}

\maketitle
\section{Introduction}
The $M$-shell ($n=3$) tungsten ions, such as S-like W$^{+58}$, are of great importance due to their potential use in plasma diagnostics in the future tokamak fusion reactor ITER~\cite{Clementson.2010.V81.p52509,Lennartsson.2013.V87.p62505,Beiersdorfer.2010.V43.p144008,Peacock.2008.V86.p277,Skinner.2009.VT134.p14022,Beiersdorfer.2015.V3.p260,Rzadkiewicz.2018.V97.p52501}. Of special interest are the many strong emission lines in the 10 $-$ 60~\AA, region, which are needed to monitor the tungsten-ion impurity levels and to properly predict the radiative emissions.

These applications stimulated some calculations of excitation energies and wavelengths for S-like W~\cite{Aggarwal.2016.V111-112.p187,Xu.2017.V95.p283,Quinet.2011.V44.p195007,Singh.2016.V49.p205002}. However, a satisfactory accuracy has not been achieved yet. For example, the two data sets reported by Aggarwal et al.~\cite{Aggarwal.2016.V111-112.p187}, using the general-purpose relativistic atomic structure package (GRASP89)~\cite{Dyall.1989.V55.p425} and the flexible atomic code (FAC)~\cite{Gu.2008.V86.p675}, are inconsistent, with excitation energy deviations of up to 30~000 cm$^{-1}$. The excitation energies calculated by Xu et al.~\cite{Xu.2017.V95.p283} differ by $3~000-70~000$~cm$^{-1}$ from the results of Ref.~\cite{Aggarwal.2016.V111-112.p187}, although both sets are evaluated using the same GRASP89 code~\cite{Dyall.1989.V55.p425}.  Unfortunately, these inconsistencies cannot be resolved by experimental measurements because the theory-observation energy deviations for both sets are much larger than the experimental error bars.

On the other hand, the identification of measured lines also needs the support of theoretical calculations, but the latter do not provide the needed accuracy.
Lennartsson et al.~\cite{Lennartsson.2013.V87.p62505} measured several lines of the $M$-shell tungsten ions using the electron beam ion trap (EBIT) facility of the Lawrence Livermore National Laboratory (LLNL). Two lines at $\lambda=34.779(4)$~{\AA} and 35.644(4)~{\AA} have not been  identified, due to the lack of robust and reliable calculations.  The FAC relativistic configuration interaction (RCI) calculations reported in \cite{Lennartsson.2013.V87.p62505} were indeed not accurate enough. To illustrate this,  the line observed at 34.779(4)~{\AA} might correspond to one of the following three transitions: $3s^23p^4$ $^1D_{2}$ $\rightarrow$ $3s^23p^4$ $^3P_{2}$ (an M1 transition of S-like W), $3s^23p^3(^2P)3d\ ^1P^o_{1} $ $\rightarrow$ $3s^23p^4$ $^3P_{2}$ (an E1 transition of S-like W), and $3s^23p^5(^2P)3d^2\ ^4D^o_{5/2}$ $\rightarrow$ $3s^23p^63d\ ^2D_{3/2}$ (an E1 transition of K-like W), with calculated wavelengths of 34.735~{\AA}, 34.800~{\AA}, and  34.812~{\AA}, respectively~\cite{Lennartsson.2013.V87.p62505}.  All three wavelengths are "equally close" to the measured one, but all lying outside the experimental error bars of 0.004~{\AA}.

The line at 35.644(4)~{\AA} measured by Lennartsson et al.~\cite{Lennartsson.2013.V87.p62505} was not  identified for the same reason.
Furthermore, many atomic energy levels of S-like W compiled in the Atomic Spectra Database (ASD) of the National Institute of Standards and Technology (NIST)~\cite{Kramida.2018.V.p}, which are determined by interpolation or extrapolation of known experimental values or by semiempirical calculation, have relatively large energy uncertainties, from 20~000 cm$^{-1}$ to 60~000 cm$^{-1}$, due to the lack of accurate theoretical reference values.

In this paper, using the multi-configuration Dirac-Hatree-Fock (MCDHF) method and the relativistic configuration interaction approach (RCI)~\citep{Fischer.2016.V49.p182004} as implemented in the GRASP2K code~\cite{Grant.2007.V.p,Joensson.2013.V184.p2197}, we improve on the accuracy of previous theoretical results.
The deviations between our wavelengths and experiments are within 0.06~\%.
The various contributions to the excitation energies, such as valence-valence (VV)
and core-valence (CV) electron correlation, along with the Breit and transverse photon interactions, are investigated in detail.
We also conduct a detailed study of the quantum electrodynamic (QED) corrections, comparing the performance of three different methods for
describing the self energy.
This effort paves the way for future applications of this approach for  accurate predictions of properties of multielectron high-$Z$ ions  and provides precision benchmarks for spectral identifications and other applications.

\section{THEORETICAL METHOD AND COMPUTATIONAL MODELS}~\label{Computational models}
\subsection{\label{subsec:calculation}Electron correlation with the MCDHF method}
In the MCDHF method~\cite{Grant.2007.V.p},
electron correlation is included by expanding the atomic state function (ASF) $\Psi\left(\Gamma PJ\right)$  in configuration state functions (CSFs)
\begin{equation}
	\Psi\left(\Gamma PJ\right)=\sum_{i=1}^Mc_i\Phi\left(\gamma_i PJ\right).
\end{equation}
The CSFs, $\Phi\left(\gamma_i PJ\right)$, are $jj$-coupled many-electron functions built from antisymmetrized products of one-electron Dirac orbitals, where
 $\gamma_i$ specifies the occupied subshells with their complete angular coupling tree information, $P$ the parity and $J$ the total angular momentum.
The radial large and small components of the one-electron orbitals and the expansion coefficients \{$ c_i $\} of the CSFs are obtained by solving iteratively the Dirac-Hartree-Fock radial equations and the configuration interaction eigenvalue problem resulting from applying the variational principle on the energy functional of the targeted states in the extended optimal level (EOL) scheme~\cite{Grant.2007.V.p,Grant.1980.V21.p207}. The energy functional is based on the  Dirac-Coulomb (DC) Hamiltonian
\begin{eqnarray}
\label{eq:DC}
\hspace{-2.0cm}
{\cal H}_{DC}
& = &
\sum_{i=1}^N \left( c \; \balpha_i \cdot {\bf p}_i
+ V_{nuc}(r_i) + c^2 (\beta_i-1)  \right)   \nonumber \\
&+& \!\sum_{j>i=1}^N \frac{1}{r_{ij}},
\end{eqnarray}
and accounts for relativistic kinematic effects.

The configurations  \{$3s^23p^4$, $3s^23p^23d^2$, $3s3p^43d$, $3s3p^23d^3$, $3p^6$, $3p^4 3d^2$\} \{$3s^23p^33d$, $3s^23p3d^3$, $3s3p^5$, $3s3p^33d^2$, $3p^53d$\} constitute, respectively, the multireference (MR) spaces for even and odd parities.
The CSF expansions are generated by allowing single (S) and double (D) excitations of all the $n = 3$  electrons, namely valence electrons, from all MR configurations to $n\leq7, l\leq5$ (i.e., up to $h$-orbital symmetry). These CSFs describe the valence-valence (VV) electron correlation.
No substitutions were allowed from the $1s$ shell, which defines an inactive closed core.
In a second series of calculations we added, to the CSFs above,  CSFs resulting from  SD-MR substitutions of all the $n = 2, 3$ electrons to $n \leq 6$, $l \leq 5$, with the restriction of allowing maximum one hole in the $n=2$ core shell. These added CSFs describe the core-valence (CV) correlation effects.
The core-core electron correlation effects are unimportant for
the excitation energies of the studied states and have thus been neglected, compare \cite{Gustafsson.2017.V5.p3}.
The numbers of CSFs distributed over the different $J$ symmetries in the final even and odd  state expansions are, respectively, 20~396~713 and 11~691~659.

\subsection{\label{subsec:Breit}Breit and QED Corrections}
In the relativistic description of the many-electron system, the Dirac-Coulomb Hamiltonian (\ref{eq:DC}) is the starting point that should be corrected by the so called transverse photon (TP) interaction, which, in the $\alpha^2$ approximation, takes the form:

\begin{eqnarray}
\label{eq:TP}
H_{\mathrm{TP}} & = & -\mathop{\sum}\limits ^N \limits_{i<j}
\left[ \frac{\boldsymbol{ \alpha}_{i} \cdot \boldsymbol{\alpha}_{j}}  {r_{ij}}  \cos(\omega_{ij}r_{ij}/c) \right. \nonumber \\
& - & \left.  (\boldsymbol{\alpha}_{i} \cdot \boldsymbol{\nabla}_{i})
 (\boldsymbol{\alpha}_{j} \cdot \boldsymbol{\nabla}_{j})\frac{\cos(\omega_{ij}r_{ij}/c)-1}{\omega^2_{ij}r_{ij}/c^2}
 \right],
\end{eqnarray}
where $\omega_{ij}$ is the frequency of the exchanged virtual photon propagating the interaction \cite{Comment1}.
In the low-frequency limit $\omega_{ij}\rightarrow 0$, the TP interaction reduces to the Breit interaction~\cite{Grant.1976.V9.p761}
\begin{equation}
\label{eq:Breit}
{\cal H}_{\mathrm{Breit}}
  =  - \sum_{j>i=1}^N \; \frac{1}{2 r_{ij}} \left[ \left(  \balpha_i \cdot \balpha_j \right) +
\frac{\left( \balpha_i \cdot {\bf r}_{ij} \right)\left( \balpha_j \cdot {\bf r}_{ij} \right)}{r^2_{ij}}\right].
\end{equation}
which is the sum of the Gaunt interaction
\begin{equation}
\label{eq:Gaunt}
{\cal H}_{\mathrm{Gaunt}} =
 - \sum_{j>i=1}^N \; \frac{    \balpha_i \cdot \balpha_j }{ r_{ij}}
\end{equation}
and the Breit retardation~ \cite{Indelicato.2007.V45.p155}
\begin{equation}
\label{eq:Breit_retardation}
{\cal H}_{\mathrm{Breit}}^{\mathrm{retard.}}
  =   + \sum_{j>i=1}^N \; \frac{1}{2 r_{ij}} \left[ \left(  \balpha_i \cdot \balpha_j \right) -
\frac{\left( \balpha_i \cdot {\bf r}_{ij} \right)\left( \balpha_j \cdot {\bf r}_{ij} \right)}{r^2_{ij}}\right].
 \end{equation}
The higher-order retardation correction beyond the Breit interaction (\ref{eq:Breit}) is therefore defined as the difference
\begin{equation}
\label{eq:HO_retardation}
{\cal H}^{\mathrm{HO}}
\equiv
{\cal H}_{\mathrm{TP}} - {\cal H}_{\mathrm{Breit}}
=
{\cal H}_{\mathrm{TP}} -
\left( {\cal H}_{\mathrm{Gaunt}} +
{\cal H}_{\mathrm{Breit}}^{\mathrm{retard.}}
\right) \; .
\end{equation}

Once the orbitals optimized through the MCDHF procedure are available, the transverse photon interaction, or the Breit interaction, and the leading QED effects (vacuum polarization and self-energy)  can be added to the Dirac-Coulomb Hamiltonian in relativistic configuration interaction (RCI) calculations to capture relativistic corrections to the Coulomb interaction.

For evaluating the TP Hamiltonian matrix elements, some decision has to be taken for the appropriate value of the $\omega_{ij}$. These matrix elements involve indeed two-body contributions  of the form
$(a_q^\dagger a_p)(a_s^\dagger a_r)$ with their own single-electron energies $\{ \epsilon_p, \epsilon_q, \epsilon_r, \epsilon_s \}$ for which $\omega_{ij}$ can be taken as $\omega_{ij} = \omega_{sr}=- \omega_{qp}$ when the effective potentials are derived ``on the energy shell''~\cite{Grant.2007.V.p}.  Averaging $\omega_{sr}$ and $\omega_{pq}$ has proved quite effective in bound state calculations involving atomic inner shells for ``off-shell'' potentials but the individual one-particle energies $\epsilon_i$ are physically meaningful only for spectroscopic orbitals. 
In the present work, the TP Hamiltonian matrix elements therefore include the frequency-dependent contributions when 
the latter involve the  spectroscopic orbitals
\[ \{ 1s_{1/2},~2s_{1/2}, ~2p_{1/2}, ~2p_{3/2}, ~3s_{1/2}, ~3p_{1/2}, ~3p_{3/2}, ~3d_{3/2}, ~3d_{5/2} \} \]
spanning the MR configurations. For contributions  involving any of the so-called correlation orbitals that are unoccupied in the MR subspace, but appear in the active orbital sets for describing electron correlation excitations,  
the low frequency limit $\omega_{ij}\rightarrow 0$ is considered.


The current status of bound state quantum electrodynamics calculations of transition energies for a few-electron highly-charged ions has been reviewed very recently by Indelicato~\cite{Indelicato.2019.V52.p232001}.
The one-electron QED corrections are separated into two contributions, namely, the self-energy (SE) and the vacuum polarization (VP). The VP
contribution  can be represented by a potential. We use for the present work the analytical expressions derived by Fullerton and Rinker ~\cite{Fullerton.1976.V13.p1283} for the Uehling model potential and the higher-order K$\ddot{\rm{a}}$ll$\acute{\rm{e}}$n-Sabry VP potential. For S-like W, the self-energy contribution dominates the QED corrections.  We investigate three different methods (M1 - M3) for estimating the latter:

\begin{itemize}
\item
QED - M1: In the current GRASP2K code, starting from the self energy of a hydrogenic system
\begin{equation}\label{SE}
\Delta E_{SE}=\left(\frac{\alpha}{\pi}\right)\frac{\alpha^2Z^4}{n^3}F(nlj,Z\alpha),
\end{equation}
where $F(nlj,Z\alpha)$ is a slowly varying function of $Z\alpha$ that has been tabulated by Mohr et al.~\cite{Mohr.1983.V29.p453} and
Klarsfeld et al.~\cite{Klarsfeld.1973.V43.p201}, the total SE contribution is given as a sum of one-electron corrections weighted by the fractional occupation number of the one-electron orbital in the total wave function. For each orbital, the effective nuclear charge or, equivalently the screening, is estimated by equating the mean radius of each MCDHF orbital to that of a hydrogenic (Dirac) orbital~~\cite{Grant.2007.V.p}.

\item
QED - M2: Starting from the latest available hydrogenic values~\cite{Mohr.1992.V45.p2727,LeBigot.2001.V64.p52508} modified to account for finite-nuclear-size effects~\cite{Mohr.1993.V70.p158,Beier.1998.V58.p954}, a  screening approximation based on the Welton interpretation~\cite{Welton.1948.V74.p1157} and implemented in GRASP2K by Lowe et al.~\cite{Lowe.2013.V85.p118}, is used to evaluate the SE contribution.

\item
QED - M3: A model QED operator, which also includes the non-local QED part to calculate the SE corrections for many-electron atomic systems, was recently developed by Shabaev et al. (QEDMOD)~\cite{Shabaev.2013.V88.p12513,Shabaev.2015.V189.p175}. We also include this model SE operator in the GRASP2K code to evaluate the SE contribution.
\end{itemize}

For all these three models,  only the contribution from the diagonal matrix elements of the QED operator
is considered. 	Further work  will quantify the off-diagonal contribution of the QED operator.~\cite{Zhang.2020.V.p}. The last two approaches (QED - M2/M3) have recently been used for investigating Breit and QED
effects in the ground-term fine structures of F-like~\cite{Li.2018.V98.p20502} and Co-like~\cite{Si.2018.V98.p012504} ions. \\

The following notations will be used for the various correlation and interaction models:
\begin{enumerate}
\item Multireference MCDHF calculations will be denoted VV when limiting the inclusion of electron correlation to the valence shells, and CV when enlarging the multiconfiguration expansions to core-valence excitations.
\item Taking the long wavelength limit for the transverse part and adding the resulting Breit interaction (\ref{eq:Breit}) to the Dirac-Coulomb Hamiltonian (\ref{eq:DC}) defines the Dirac-Coulomb-Breit (DCB) Hamiltonian, in the effective Coulomb gauge \begin{equation}
\label{eq:DCB}
{\cal H}_{DCB}
= {\cal H}_{DC} + {\cal H}_{Breit} \; .
\end{equation}
\item Adding the transverse photon interaction in Coulomb gauge (\ref{eq:TP}) to the Dirac-Coulomb Hamiltonian (\ref{eq:DC}) gives the more complete Hamiltonian
\begin{equation}
\label{eq:DCTP}
{\cal H}_{DCTP}
= {\cal H}_{DC} +  {\cal H}_{TP}
\end{equation}
\item Calculations including QED corrections estimated by selecting one of the three models (M1, M2 or M3) as described above, and added to ${\cal H}_{DCTP}$ in the very last step, are denoted QED(Mx).
\end{enumerate}

The relativistic corrections to the Coulomb interaction and quantum electrodynamics  corrections considered in steps (2-4) are included in
RCI calculation based on CSF expansions accounting for both
VV and CV electron correlation.

\section{\label{sec:results and discussion}Results}
\subsection{Excitation energies}~\label{transition}
\subsubsection{Electron correlation}
In Table~\ref{Table1}, we present the excitation energies for a selection of levels from the above correlation and interaction models, together with the values compiled in the NIST Atomic Spectra Database (ASD)~\cite{Kramida.2018.V.p}. Only the levels, for which  the NIST compiled values deduced from measured lines are listed in Table~\ref{Table1}. The atomic units are used throughout the present work, if units are not indicated explicitly.
The deviations ($\Delta E=E_{\rm{MCDHF}/\rm{RCI}}-E_{\rm{NIST}}$) between our calculated MCDHF/RCI excitation energies and the experimental values compiled in the NIST ASD are also reported.
On average, CV electron correlation plays a smaller role than VV electron correlation, as expected for transitions involving valence excitations. CV electron correlation was systematically omitted in all previous theoretical calculations performed for S-like W~\cite{Fullerton.1976.V13.p1283,Aggarwal.2016.V111-112.p187,Singh.2016.V49.p205002,Xu.2017.V95.p283}. However, limiting electron correlation to VV electron correlation  is not enough to reach the needed accuracy for assisting   spectroscopists in the spectral lines identification process, as discussed in Ref.~\cite{Lennartsson.2013.V87.p62505}.
By comparing the two columns
 $\Delta E$-VV and  $\Delta E$-CV of Table~\ref{Table1}, it is seen that the addition of CV to the VV electron correlation further reduces the energy differences between the MCDHF and observed (NIST) excitation energy values by $\simeq 1~100-5~600$~cm$^{-1}$ for the levels considered. This illustrates the importance of core-valence correlation, even for such highly charged ions.

\subsubsection{The Breit interaction and QED corrections}
As revealed by column~5 of  Table~\ref{Table1}, the magnitude of the Breit correction to excitation energies strongly depends on the electronic configuration.
The Breit correction   affects the  excitation energies of the levels
of the $3s^2 3p^3 3d$ configuration by $\simeq 10~000 - 45~000$~cm$^{-1}$.  The corresponding
effect on the levels arising from the $3s 3p^5$ configuration is considerably smaller, around $5~000$~cm$^{-1}$.

Comparing column~6 with column~5 of  Table~\ref{Table1}, one observes that  the higher-order frequency-dependent corrections
${\cal H}^{\mathrm{HO}} = {\cal H}_{\mathrm{TP}} - {\cal H}_{\mathrm{Breit}} $
are relatively small compared with the Breit interaction, but cannot be neglected for precision calculations of excitation energies in S-like W.

Adding the QED corrections to the MCDHF/RCI  excitation energies improves substantially the agreement with observation. These QED corrections reach around  4~000 - 29~000~cm$^{-1}$.
The QED corrections to excitation energies are naturally grouped according to the electronic configuration of the level considered, as observed for the Breit interaction.
As expected the QED contribution to the excitation energies of the levels of the $3s 3p^5$ configuration is significantly larger than the contribution
to the  excitation energies of the levels of the $3s^2 3p^3 3d$ configuration
due to the change in the $3s$ electron occupation number for the former configuration.

The QED corrections to the excitation energies obtained using the three different QED potentials as described above, are also given in  Table~\ref{Table1}.  Compared with the results obtained by using the QED-M1 method,  the MCDHF/RCI results based on the QED - M2 and QED - M3 methods are closer to the experimental NIST values.  The QED corrections obtained by using the QED - M2 and QED - M3 methods are very similar. For each level considered, the excitation energies obtained with both QED-M2/M3 models lie within the error bars of the estimated experimental uncertainty  reported in the $E_{\rm NIST}$ column.

In short, the Breit interaction and QED corrections play the most  important role in the calculations of excitation energies of S-like W$^{58+}$.
However, the CV electron correlation and the higher-order corrections (\ref{eq:HO_retardation}) arising from
${\cal H}_{\mathrm{TP}} - {\cal H}_{\mathrm{Breit}}$, which were not considered in the previous calculations~\cite{Quinet.2011.V44.p195007,Aggarwal.2016.V111-112.p187,Singh.2016.V49.p205002,Xu.2017.V95.p283}, cannot be omitted for high-precision results. Since our MCDHF/RCI excitation energies obtained by using the QED - M2 and QED - M3 methods are very similar, the results that are reported
in the following sections are only based on the M2 model. Moreover, the MCDHF/RCI label will be shortened from here by the single MCDHF generic denomination to simplify the notations, but the reader should be aware that Breit, TP and QED corrections were all included in the final RCI calculations.

\subsection{Wavelengths and transition rates}~\label{transition}
In Table~\ref{Table2} we present the  differences $\Delta \lambda$ between the present theoretical wavelengths calculated at different levels
of approximation and the experimental values.
The differences between the present theoretical wavelengths  and the measured values for the E1 transitions are found to be around several hundreds m{\AA} when  VV and CV electron correlation is included in the Dirac-Coulomb approximation.  Once the ${\cal H}_{DCTP}$ Hamiltonian (\ref{eq:DCTP}) is  considered to take the transverse photon interaction into account, our  wavelengths   are getting closer to the measured ones, reducing the
differences to $-200 \le \Delta \lambda \le +24$~m{\AA}. By further adding the QED corrections, the  wavelength differences  become of the same order of magnitude than the estimated uncertainty of the experimental value reported in parentheses in the $\lambda_{\rm{exp.}}$ column.  Since the upper and lower levels of the M1 transition $3s^2 3p^4~^1D_{2}	-	3s^23p^4~^3P_{2}$  belong to the same configuration, the Breit/TP interaction and QED corrections have similar effects on the levels involved, affecting only slightly the wavelength of this intra-configuration transition.

As far as transition rates are concerned, the magnetic Breit/TP interaction decreases transition rates by $\simeq 3$~\%, on average. However, the variations in transition rates due to QED
are about $\simeq 0.3$~\%. One observes that the QED corrections barely change the M1 transition rate. This characteristics was also found for the M1 transitions within the $3d^n$ configurations (with $n=2-5$) in Ref.~\cite{Safronova.2018.V97.p12502}.

\subsection{Comparison with other theoretical works and observation}~\label{Results}
Excitation energies of S-like W from the present MCDHF/RCI calculations, as well as the compiled data from the NIST ASD~\cite{Kramida.2018.V.p}, are listed in Table~\ref{Table3}.
For comparison, the two theoretical data sets reported by Aggarwal et al.~\cite{Aggarwal.2016.V111-112.p187},  and the theoretical results provided by Xu et al.~\cite{Xu.2017.V95.p283} are also included in the table.

The NIST compiled values in square brackets are determined from semi-empirical calculations by Kramida~\cite{Kramida.2011.V89.p551} using Cowan's code~\cite{Cowan.1981.V.p}. The other NIST values are deduced from measured lines that were observed using the EBIT facilities~\cite{Ralchenko.2008.V41.p21003,Clementson.2010.V81.p52509}. For each level, the number reported in parenthesis, after the NIST excitation level energy, is the estimated accuracy provided by the NIST ASD. It can be seen from this table that the accuracy of the NIST values quoted in square brackets is generally about tens of thousands cm$^{-1}$, whereas the NIST values deduced by measured lines are much more accurate ($110 - 2500$~ cm$^{-1}$).

The  energy differences, $\Delta E=E_{\rm{theory}}-E_{\rm{NIST}}$, between the different theoretical excitation energies (MCDHF, Aggarwal1, Aggarwal2, and Xu) and the NIST compiled values  are also reported in Table~\ref{Table3}. The differences $\Delta E$ between the present MCDHF/RCI energies and the NIST values deduced from measured lines are well controlled within 2~800 cm$^{-1}$, and are generally within or smaller than  the NIST estimated uncertainties. On the contrary, the two theoretical data sets  of Aggarwal et al.~\cite{Aggarwal.2016.V111-112.p187} deviate from the NIST measured values by up to $\simeq 15~700$~ cm$^{-1}$. Moreover, these two data sets do not support each other well, revealing deviations of up to $\simeq 11~000$~ cm$^{-1}$. Similarly, the excitation energies calculated by Xu et al.~\cite{Xu.2017.V95.p283} differ from the NIST measured values by up to 15~000~cm$^{-1}$.

The differences between the previous calculations of atomic energy levels~\cite{Aggarwal.2016.V111-112.p187,Xu.2017.V95.p283} and the NIST measured value are several times or one order of magnitude larger than the corresponding differences calculated for the present theoretical MCDHF/RCI energies. This indicates that the present theoretical excitation energies of S-like W are highly accurate and represent a great improvement on the latest theoretical results~\cite{Aggarwal.2016.V111-112.p187,Xu.2017.V95.p283}. In addition, excitation energies  in Table~\ref{Table3} are presented in the order of the present theoretical excitation energies. The results from the previous calculations~\cite{Aggarwal.2016.V111-112.p187,Xu.2017.V95.p283} that do not correspond to this order are highlighted in boldface. This explicitly illustrates that the order of the levels from the previous calculations  is not always correct, although some levels are very close to each other, in which case the order remains uncertain.

Looking at the NIST values that are reported in square brackets to mark their origin from semi-empirical parametric calculations, their differences with the present theoretical values are usually about tens of thousands cm$^{-1}$, with the largest difference of 46~000~cm$^{-1}$. For this reason, our MCDHF/RCI excitation energies, compared with these NIST compiled values, also represent a substantial improvement in accuracy. We therefore recommend the use of the  present theoretical values for updating these NIST semi-empirical data,  and even suggest their use as input data for a new parametric fit using Cowan's code, which would increase dramatically the accuracy and quality of the NIST compiled values.

Spectroscopists pay close attention to the $n = 3 \rightarrow n = 3$ transitions of S-like W that can be used as benchmarks for advancing electron-correlation physics in multi-electron high-$Z$ ions. Furthermore, the $n = 3 \rightarrow n = 3$ forbidden transitions, such as $3p - 3p$ and $3d - 3d$,  are also important for plasma diagnostics because their line intensity ratios are highly sensitive to the electron density. We compare in Table~\ref{Table4} the  present MCDHF/RCI wavelengths with the measured values in the range of 10 {\AA} to 60 {\AA}~\cite{Ralchenko.2008.V41.p21003,Lennartsson.2013.V87.p62505,Clementson.2010.V81.p52509}, as well as with previous theoretical values (Aggarwal1,  Aggarwal2, and Xu )~\cite{Aggarwal.2016.V111-112.p187,Xu.2017.V95.p283} . The theory-observation deviations $\Delta \lambda$ (in m{\AA})  are also listed in the same table. The agreement between the experimental and present theoretical wavelength values is generally within 10~m{\AA} for the transitions in the X-ray region. This signifies that the accuracy of our calculations is high enough to confirm or revise experimental identifications. For comparison, the results from Ref.~\cite{Aggarwal.2016.V111-112.p187} and from Ref.~\cite{Xu.2017.V95.p283} deviate from the measured values by up to 130 m{\AA} and 336 m{\AA}, respectively. Their differences with the experimental wavelength values are also usually several times or one order of magnitude larger compared with the corresponding MCDHF/RCI differences.

The line at 34.779(4)~{\AA}, measured by Lennartsson et al.~\cite{Lennartsson.2013.V87.p62505} using the  EBIT facility, was not explicitly identified, since relatively limited  RCI calculations were available for supporting line assignments. The calculated RCI values for the $3s^23p^4$ $^1D_{2}$ $\rightarrow$ $3s^23p^4$ $^3P_{2}$ (an M1 transition) and $3s^23p^3(^2P)3d\ ^1P_{1} $ $\rightarrow$ $3s^23p^4$ $^3P_{2}$ (an E1 transition) in Ref.~\cite{Lennartsson.2013.V87.p62505}, are respectively 34.735~{\AA}, 34.800~{\AA}. They are "equally close" to the measured wavelength of 34.779(4)~{\AA}. By comparison, our  MCDHF values are, respectively, 34.819~{\AA} and 34.773~{\AA} for these E1 and M1 transitions. Our theoretical wavelength $\lambda= 34.773$~{\AA} for the M1 transition agrees well enough with the measured wavelength at $ \lambda$=34.779(4)~{\AA} to suggest to assign the latter to the M1 transition, but not to the E1 transition.

Among the previous different calculations~\cite{Aggarwal.2016.V111-112.p187,Xu.2017.V95.p283}, the M1 transitions are not reported in Ref.~\cite{Xu.2017.V95.p283}. Therefore, this theoretical work cannot be used to assign the line 34.779(4)~{\AA} due to incomplete data. The results provided in Ref.~\cite{Aggarwal.2016.V111-112.p187} for these E1 and M1 transitions are, respectively, 34.74~{\AA} and 34.78~{\AA}. By comparison, the present MCDHF/RCI values are respectively 34.819~{\AA} and 34.773~{\AA}, i.e. a wavelength for the E1 transition longer than for the M1 transition. This fact alone illustrates that the order of the $3s^23p^4$ $^1D_{2}$ and $3s^23p^3(^2P)3d\ ^1P_{1}$ levels found in the calculations~\cite{Aggarwal.2016.V111-112.p187} is most likely not correct, as pointed out above.

Accurate wavelengths ($\lambda$), transition rates ($A$), weighted oscillator strengths ($gf$) and line strengths ($S$) for E1, E2, M1 and M2 transitions with a radiative branching ratio larger than 0.1~\% involving the lowest 88 levels from the present MCDHF/RCI calculations are listed in Table~\ref{Table5}. All E1 and E2 transitions are calculated in Babushkin (length) gauge.  Compared with the calculations~\cite{Aggarwal.2016.V111-112.p187}, the present theoretical calculations also provide a complete  data set of accurate radiative transition data. Aggarwal et al. stated that calculations were performed for the transitions among the lowest 220 levels of the  $n=3$ configurations, whereas  radiative rates were only reported for the transitions involving the two lowest levels (The data involving the higher levels did not belong to S-like W). Future modeling and diagnosing of plasmas would benefit from the present complete data sets of high accuracy. The present work could also be used for cross-checking  work under progress \cite{Piibeleht.2019.V.p} on the inclusion of QED corrections in GRASP2018~\cite{FroeseFischer.2019.V237.p184}.

\section{Conclusion}~\label{conclusion}
We calculated the energy levels, wavelengths, and E1, E2, M1, and M2 transition parameters among the 88 lowest levels for S-like W  using the MCDHF and RCI methods~\citep{Grant.2007.V.p} implemented in the GRASP2K package~\cite{Grant.2007.V.p,Joensson.2013.V184.p2197}.
We analyzed in detail the relative importance of different physical effects, namely, VV and CV electron correlations, the Breit interaction, the higher-order frequency-dependent retardation correction through the Transverse Photon interaction, and the QED corrections, using for the latter three different models.

The Breit and QED corrections  play an important role in the calculations of  excitation energies and wavelengths in S-like~W. The CV electron and the higher-order retardation corrections beyond the Breit interaction, which were not considered in previous calculations~\cite{Quinet.2011.V44.p195007,Aggarwal.2016.V111-112.p187,Singh.2016.V49.p205002,Xu.2017.V95.p283}, should not be ignored for getting high-precision results. The present set of results is accurate enough to support and help spectroscopists in their delicate and challenging task of spectral lines identification. We expect that the present complete and accurate atomic data set for S-like W would benefit future modeling and diagnosing of plasmas.

\section{ACKNOWLEDGMENTS}
We acknowledge the support from the National Key Research and Development Program of China under Grant No. 2017YFA0402300, the National Natural Science Foundation of China (Grants No. 11703004, No. 11674066 and No. 11974080), the Natural Science Foundation of Hebei Province, China (A2019201300 and A2017201165), and the Natural Science Foundation of Educational Department of Hebei Province, China (BJ2018058). This work is also supported by the Fonds de la Recherche Scientifique - (FNRS) and the Fonds Wetenschappelijk Onderzoek - Vlaanderen (FWO) under EOS Project n$^{\rm o}$~O022818F, and by the Swedish research council under contracts 2015-04842 and 2016-04185. K.W. expresses his gratitude to the support from the visiting researcher program at the Fudan University.

\linespread{1}
\bibliographystyle{aip}
\bibliography{reference.bib}

\begin{thebibliography}{10}

\bibitem{Clementson.2010.V81.p52509}
J.~Clementson and P.~Beiersdorfer,
\newblock Phys. Rev. A {\bf 81}, 052509 (2010).

\bibitem{Lennartsson.2013.V87.p62505}
T.~Lennartsson, J.~Clementson, and P.~Beiersdorfer,
\newblock Phys. Rev. A {\bf 87}, 062505 (2013).

\bibitem{Beiersdorfer.2010.V43.p144008}
P.~Beiersdorfer et~al.,
\newblock J. Phys. B: At., Mol. Opt. Phys. {\bf 43}, 144008 (2010).

\bibitem{Peacock.2008.V86.p277}
N.~J. Peacock, M.~G. O'Mullane, R.~Barnsley, and M.~Tarbutt,
\newblock Can. J. Phys. {\bf 86}, 277 (2008).

\bibitem{Skinner.2009.VT134.p14022}
C.~H. Skinner,
\newblock Phys. Scr. {\bf T134}, 014022 (2009).

\bibitem{Beiersdorfer.2015.V3.p260}
P.~Beiersdorfer, J.~Clementson, and U.~I. Safronova,
\newblock Atoms {\bf 3}, 260 (2015).

\bibitem{Rzadkiewicz.2018.V97.p52501}
J.~Rzadkiewicz et~al.,
\newblock Phys. Rev. A {\bf 97}, 052501 (2018).

\bibitem{Aggarwal.2016.V111-112.p187}
K.~M. Aggarwal and F.~P. Keenan,
\newblock At. Data Nucl. Data Tables {\bf 111-112}, 187  (2016).

\bibitem{Xu.2017.V95.p283}
M.~Xu, A.~Yan, S.~Wu, F.~Hu, and X.~Li,
\newblock Can. J. Phys. {\bf 95}, 283 (2017).

\bibitem{Quinet.2011.V44.p195007}
P.~Quinet,
\newblock J. Phys. B: At. Mol. Opt. Phys {\bf 44}, 195007 (2011).

\bibitem{Singh.2016.V49.p205002}
G.~Singh and N.~K. Puri,
\newblock J. Phys. B: At., Mol. Opt. Phys. {\bf 49}, 205002 (2016).

\bibitem{Dyall.1989.V55.p425}
K.~G. Dyall, I.~P. Grant, C.~T. Johnson, F.~A. Parpia, and E.~P. Plummer,
\newblock Comput. Phys. Commun. {\bf 55}, 425 (1989).

\bibitem{Gu.2008.V86.p675}
M.~F. Gu,
\newblock Can. J. Phys {\bf 86}, 675 (2008).

\bibitem{Kramida.2018.V.p}
A.~Kramida, {Yu.~Ralchenko}, J.~Reader, and {and NIST ASD Team},
\newblock {NIST Atomic Spectra Database (ver. 5.5.3), [Online]. Available:
  {\tt{https://physics.nist.gov/asd}} [2018, March 27]. National Institute of
  Standards and Technology, Gaithersburg, MD.}, 2018.

\bibitem{Fischer.2016.V49.p182004}
C.~{Froese Fischer}, M.~Godefroid, T.~Brage, P.~J$\ddot{\rm{o}}$nsson, and
  G.~Gaigalas,
\newblock J. Phys. B: At., Mol. Opt. Phys. {\bf 49}, 182004 (2016).

\bibitem{Grant.2007.V.p}
I.~P. Grant,
\newblock Springer Sci. +Bus. Media, LLC, New York, NY, 2007.

\bibitem{Joensson.2013.V184.p2197}
P.~J$\ddot{\rm{o}}$nsson, G.~Gaigalas, J.~Biero$\acute{\rm{n}}$, C.~{Froese
  Fischer}, and I.~P. Grant,
\newblock Comput. Phys. Commun. {\bf 184}, 2197 (2013).

\bibitem{Grant.1980.V21.p207}
I.~P. Grant, B.~J. McKenzie, P.~H. Norrington, D.~F. Mayers, and N.~C. Pyper,
\newblock Comput. Phys. Commun. {\bf 21}, 207  (1980).

\bibitem{Gustafsson.2017.V5.p3}
S.~Gustafsson, P.~J$\ddot{\rm{o}}$nsson, C.~{Froese Fischer}, and I.~P. Grant,
\newblock Atoms {\bf 5}, 3 (2017).

\bibitem{Comment1}
We would like to stress the importance of the $-$ sign of the second term of
  (3) inside the brackets. That term is indeed often wrongly written with a $+$
  sign (see i.e.~\cite{Si.2018.V98.p012504}). Referring to the original paper
  by Grant and Pyper \cite{Grant.1976.V9.p761}, the $+$ sign can be used when
  omitting subscripts $i$ and $j$ of the gradient operators, i.e. involving
  differentiation with respect to ${\bm r}_{ij} = {\bm r}_i - {\bm r}_j$ as
  correctly written in \cite{Fischer.2016.V49.p182004}.

\bibitem{Grant.1976.V9.p761}
I.~P. Grant and N.~C. Pyper,
\newblock J. Phys. B: At. Mol. Opt. Phys. {\bf 9}, 761 (1976).

\bibitem{Indelicato.2007.V45.p155}
P.~Indelicato, J.~P. Santos, S.~Boucard, and J.-P. Desclaux,
\newblock Eur. Phys. J. D {\bf 45}, 155 (2007).

\bibitem{Indelicato.2019.V52.p232001}
P.~Indelicato,
\newblock J. Phys. B: At. Mol. Opt. Phys {\bf 52}, 232001 (2019).

\bibitem{Fullerton.1976.V13.p1283}
L.~W. Fullerton and G.~A. Rinker,
\newblock Phys. Rev. A {\bf 13}, 1283 (1976).

\bibitem{Mohr.1983.V29.p453}
P.~J. Mohr,
\newblock At. Data Nucl. Data Tables {\bf 29}, 453  (1983).

\bibitem{Klarsfeld.1973.V43.p201}
S.~Klarsfeld and A.~Maquet,
\newblock Phys. Lett. B {\bf 43}, 201  (1973).

\bibitem{Mohr.1992.V45.p2727}
P.~J. Mohr and Y.~K. Kim,
\newblock Phys. Rev. A {\bf 45}, 2727 (1992).

\bibitem{LeBigot.2001.V64.p52508}
E.~O. Le~Bigot, P.~Indelicato, and P.~J. Mohr,
\newblock Phys. Rev. A {\bf 64}, 052508 (2001).

\bibitem{Mohr.1993.V70.p158}
P.~J. Mohr and G.~Soff,
\newblock Phys. Rev. Lett. {\bf 70}, 158 (1993).

\bibitem{Beier.1998.V58.p954}
T.~Beier, P.~J. Mohr, H.~Persson, and G.~Soff,
\newblock Phys. Rev. A {\bf 58}, 954 (1998).

\bibitem{Welton.1948.V74.p1157}
T.~A. Welton,
\newblock Phys. Rev. {\bf 74}, 1157 (1948).

\bibitem{Lowe.2013.V85.p118}
J.~A. Lowe, C.~T. Chantler, and I.~P. Grant,
\newblock Radiat. Phys. Chem. {\bf 85}, 118  (2013).

\bibitem{Shabaev.2013.V88.p12513}
V.~M. Shabaev, I.~I. Tupitsyn, and V.~A. Yerokhin,
\newblock Phys. Rev. A {\bf 88}, 012513 (2013).

\bibitem{Shabaev.2015.V189.p175}
V.~M. Shabaev, I.~I. Tupitsyn, and V.~A. Yerokhin,
\newblock Comput. Phys. Commun. {\bf 189}, 175  (2015).

\bibitem{Zhang.2020.V.p}
C.~Y. Zhang et~al.,
\newblock work in progress.

\bibitem{Li.2018.V98.p20502}
M.~C. Li, R.~Si, T.~Brage, R.~Hutton, and Y.~M. Zou,
\newblock Phys. Rev. A {\bf 98}, 020502 (2018).

\bibitem{Si.2018.V98.p012504}
R.~Si et~al.,
\newblock Phys. Rev. A {\bf 98}, 012504 (2018).

\bibitem{Safronova.2018.V97.p12502}
M.~S. Safronova, U.~I. Safronova, S.~G. Porsev, M.~G. Kozlov, and
  {Yu.~Ralchenko},
\newblock Phys. Rev. A {\bf 97}, 012502 (2018).

\bibitem{Kramida.2011.V89.p551}
A.~Kramida,
\newblock Can. J. Phys. {\bf 89}, 551 (2011).

\bibitem{Cowan.1981.V.p}
R.~D. Cowan,
\newblock Berkeley, CA: Univ. California Press, 1981.

\bibitem{Ralchenko.2008.V41.p21003}
{Yu.~Ralchenko} et~al.,
\newblock J. Phys. B: At. Mol. Opt. Phys. {\bf 41}, 021003 (2008).

\bibitem{Piibeleht.2019.V.p}
M.~Piibeleht, L.~Pa\u{s}teka, R.~Si, J.~Grumer, and P.~Schwerdtfeger,
\newblock {New QED operators and updated CI program in GRASP}, 2019,
\newblock Poster presented at the Solvay Workshop on "New Frontiers in Atomic,
  Nuclear, Plasma and Astrophysics", Brussels, November25-27.

\bibitem{FroeseFischer.2019.V237.p184}
C.~{Froese Fischer}, G.~Gaigalas, P.~J$\ddot{\rm{o}}$nsson, and
  J.~Biero$\acute{\rm{n}}$,
\newblock Comput. Phys. Commun. {\bf 237}, 184  (2019).

\end{thebibliography}

\clearpage
\setlength{\tabcolsep}{3pt}

\tiny

\begin{table*}
	\tiny
	\caption{ \label{Table1}
		Excitation energies ($E$, in cm$^{-1}$) from the present MCDHF/RCI calculations,  compared with experimental values compiled in the NIST ASD~\cite{Kramida.2018.V.p}. The estimated uncertainty of the experimental value for each level is reported in brackets in the $E_{\rm NIST}$ column. The MCDHF values were calculated using the Dirac-Coulomb Hamitonian with CSF expansions targeting valence (VV) and  core-valence (CV) electron correlation; The DCB and DCTP values were obtained by considering in the RCI step the Dirac-Coulomb-Breit (\ref{eq:DCB}) and Dirac-Coulomb-Transverse-Photon (\ref{eq:DCTP}) Hamiltonians, respectively. The QED corrections were estimated using the three different models (M1-M3)(see the text for details). The differences $\Delta E=E_{\rm{MCDHF/RCI}}-E_{\rm{NIST}}$  are also reported (in cm$^{-1}$). The key in the first column is a number assigned to each level, and is available in the Table~\ref{Table3}.}
	\begin{tabular}{ccccccccccccccccccc}
		&& & & & & & & &  &&&&&&  &  & & \\
		\colrule 		\colrule
		&& & & & & & & &  &&&&&&  &  & & \\
		Key &\multirow{6}{*}{Level}&\multirow{7}{*}{$E_{\rm NIST}$ (cm$^{-1}$)} & \multicolumn{2}{c}{$E_{\rm MCDHF}$ (cm$^{-1}$)} &
		\multicolumn{5}{c}{$E_{\rm RCI}$ (cm$^{-1}$)} &&&
		\multicolumn{7}{c}{$
			\Delta E = E_{\rm{MCDHF/RCI}}-E_{\rm{NIST}}$ (in cm$^{-1}$)} \\
		&& & & & & & & &  &&&&&&  &  & & \\
		\cline{4-5} \cline{7-11}   \cline{13-19} \\
		&& & VV  & $+$CV && DCB & DCTP & \multicolumn{3}{c}{$+$QED} &&VV  &  $+$CV & DCB & DCTP & \multicolumn{3}{c}{$+$QED}  \\
		&& & & & & & & &  &&&&&&  &  & & \\
		\cline{9-11} \cline{17-19}\\
		& & & & & & & & M1 & M2 & M3 & & & & & & M1 & M2 & M3 \\
		& & & & & & & & &  &&&&&&  &  & & \\
		\colrule
		& & & & & & & &    &    &    & & & & &    &    &  & \\
		3&	$	   3s^23p^3(^2P)3d\ ^3D_{2}    	^o	$	&	1893900(700)	&	1908452 	&	1906355 	&&	1896235 	&	1900096 	&	1894013 	&	1893744 	&	1893519 	&		&	14552	&	12455	&	2335	&	6196 	&	113	&	-156 	&	-381	\\
		6&	$	  3s^23p^3(^2P)3d\ ^3F_{3}  	^o	$	&	1966200(800)	&	1985487 	&	1983233 	&&	1968900 	&	1972761 	&	1966720 	&	1966449 	&	1966226 	&		&	19287	&	17033	&	2700	&	6561 	&	520	&	249 	&	26	\\
		7&	$	    3s^23p^3(^2P)3d\ ^3F_{4}    	^o	$	&	2574320(110)	&	2608264 	&	2606644 	&&	2575895 	&	2579342 	&	2574485 	&	2574205 	&	2574094 	&		&	33944	&	32324	&	1575	&	5022 	&	165	&	-115 	&	-226	\\
		15&	$	   3s3p^5\ ^3P_{2}    	^o	$	&	4282700(1800)	&	4322043 	&	4320956 	&&	4315574 	&	4311808 	&	4281212 	&	4282365 	&	4282846 	&		&	39343	&	38256	&	32874	&	29108 	&	-1488	&	-335 	&	146	\\
		27&	$	3s^2 3p^3(^2D)3d\ ^1F_{3}  	^o	$	&	4963500(1500)	&	5017897 	&	5013815 	&&	4968599 	&	4968098 	&	4964990 	&	4964223 	&	4964236 	&		&	54397	&	50315	&	5099	&	4598 	&	1490	&	723 	&	736	\\
		28&	$	3s^2 3p^3(^2D)3d\ ^3P_{2}    	^o	$	&	5018300(2500)	&	5071977 	&	5066689 	&&	5028431 	&	5027550 	&	5021539 	&	5020965 	&	5021025 	&		&	53677	&	48389	&	10131	&	9250 	&	3239	&	2665 	&	2725	\\
		29&	$	3s^2 3p^3(^2D)3d\ ^3S_{1}  	^o	$	&	5062800(2100)	&	5114885 	&	5109250 	&&	5074042 	&	5073037 	&	5066102 	&	5065599 	&	5065679 	&		&	52085	&	46450	&	11242	&	10237 	&	3302	&	2799 	&	2879	\\
		&&&&&& &  &  &&&&&&  &  & \\
		\colrule 	\colrule
	\end{tabular}
\end{table*}
\clearpage
\setlength{\tabcolsep}{3pt}
\scriptsize

\begin{longtable*}{@{\extracolsep\fill}ccccccccccccc@{}}
	\caption{Wavelength deviations $\Delta \lambda$ (in m{\AA}) between the present theoretical $\lambda$-values calculated in different levels of approximations and the measured values $\lambda_{\rm{exp.}}$  (in {\AA}) together with theoretical transition rates $A$ (in s$^{-1}$). The third column specifies the transition mode (TM) considered for the reported rate. For each transition, the estimated uncertainty in the experimental wavelength value  is reported in brackets in the $\lambda_{\rm{exp.}}$ column.
		Valence and core-valence electron correlation are included through the MCDHF calculations using the Dirac-Coulomb (DC) Hamiltonian.
		The DCTP value result from the RCI calculations including the transverse photon interaction, as described in the text. The $+$QED columns report values obtained by adding the QED(M2) corrections to the DCTP Hamiltonian. } \label{Table2}\\
	\hline\hline\\
	\multirow{2}{*}{Upper level} & \multirow{2}{*}{Lower level} & \multirow{2}{*}{TM} & \multirow{2}{*}{$\lambda_{\rm{exp.}}$  ({\AA})} &
	\multicolumn{3}{c}{$\Delta \lambda$ (m{\AA}) }  & & \multicolumn{3}{c}{$A$ (s$^{-1}$)}\\
	\cline{5-7}  \cline{9-11} \\
	
	& &   & &  DC & DCTP & $+$QED & &   DC & DCTP & $+$QED \\
	
	\hline\\
	\endhead
	
	$	3s^23p^3(^2D)3d$ $^3S_{1}	^o	$	&	$	3s^23p^4$ $^3P_{2}	$	&	 E1  	 &	$	19.752(8)^a	$	 &	-180	 &	-40	  &	-11	&	&	$	4.838\rm{E}+12	$	&	$	4.672\rm{E}+12	$	&	$	4.737\rm{E}+12	$	\\
	$	3s^23p^3(^2D)3d$ $^3P_{2}	^o	$	&	$	3s^23p^4$ $^3P_{2}	$	&	E1	&	$	19.927(10)^a	$	 &	-227	 &	-27	  &	-11	&	&	$	4.201\rm{E}+12	$	&	$	4.176\rm{E}+12	$	&	$	4.095\rm{E}+12	$	\\
	$	3s^23p^3(^2D)3d$ $^1F_{3}	^o	$	&	$	3s^23p^4$ $^3P_{2}	$	&	E1	&	$	20.147(6)^a	$	 &	-247	 &	-47	  &	-3	&	&	$	2.981\rm{E}+12	$	&	$	2.902\rm{E}+12	$	&	$	2.894\rm{E}+12	$	\\
	$	3s3p^5$ $^3P_{2}	^o	$	&	$	3s^23p^4$ $^3P_{2}	$	&	E1	&	$	23.350(1)^a	$	 &	-250	 &	-150	  &	2	&	&	$	8.616\rm{E}+11	$	&	$	8.090\rm{E}+11	$	&	$	8.314\rm{E}+11	$	\\
	$	3s^23p^4$ $^1D_{2}		$	&	$	3s^23p^4$ $^3P_{2}	$	&	M1	&	$	34.779(4)^b	$	 &	7	 &	24	  &	-6	&	&	$	2.076\rm{E}+08	$	&	$	1.999\rm{E}+08	$	&	$	2.004\rm{E}+08	$	\\
	$	3s^23p^3(^2P)3d$ $^3D_{3}	^o	$	&	$	3s^23p^4$ $^3P_{2}	$	&	E1	&	$	35.974(2)^b	$	 &	-374	 &	-74	  &	-12	&	&	$	5.293\rm{E}+11	$	&	$	5.143\rm{E}+11	$	&	$	5.112\rm{E}+11	$	\\
	$	3s^23p^3(^2P)3d$ $^1P_{1}	^o	$	&	$	3s^23p^4$ $^1S_{0}	$	&	E1	&	$	36.881(3)^b	$	 &	-381	 &	-81	  &	-11	&	&	$	2.814\rm{E}+11	$	&	$	2.735\rm{E}+11	$	&	$	2.715\rm{E}+11	$	\\
	$	3s^23p^3(^2P)3d$ $^3P_{2}	^o	$	&	$	3s^23p^4$ $^3P_{2}	$	&	E1	&	$	38.072(2)^b	$	 &	-472	 &	-72	  &	-10	&	&	$	1.486\rm{E}+11	$	&	$	1.448\rm{E}+11	$	&	$	1.431\rm{E}+11	$	\\
	$	3s^23p^3(^2P)3d$ $^3F_{3}	^o	$	&	$	3s^23p^4$ $^3P_{2}	$	&	E1	&	$	50.86(2)^c	$	 &	-460	 &	-160	  &	-7	&	&	$	1.741\rm{E}+09	$	&	$	1.626\rm{E}+09	$	&	$	1.616\rm{E}+09	$	\\
	$	3s^23p^3(^2P)3d$ $^3D_{2}	^o	$	&	$	3s^23p^4$ $^3P_{2}	$	&	E1	&	$	52.80(2))^c	$	 &	-300	 &	-200	  &	5	&	&	$	9.009\rm{E}+09	$	&	$	8.883\rm{E}+09	$	&	$	8.784\rm{E}+09	$	\\

	\hline
\end{longtable*}

\noindent
$^a$ From Ralchenko et al.~\cite{Ralchenko.2008.V41.p21003} \\
$^b$ From Lennartsson et al.~\cite{Lennartsson.2013.V87.p62505} \\
$^c$ From Clementson et al.~\cite{Clementson.2010.V81.p52509} \\

\clearpage
\clearpage
\setlength{\tabcolsep}{0.5\tabcolsep}
\scriptsize

\begin{longtable*}{@{\extracolsep\fill}lcccccccccccc@{}}
	\caption{Computed excitation energies ($E$, in cm$^{-1}$) from the present MCDHF/RCI calculations (DCTP Hamiltonian + QED-M2), as well as  from the previous theoretical works~\cite{Aggarwal.2016.V111-112.p187} (Aggarwal1 and Aggarwal2) and~\cite{Xu.2017.V95.p283}~(Xu), are compared with the values compiled in the NIST ASD~\cite{Kramida.2018.V.p}.  The differences ($\Delta E$, in cm$^{-1}$) of the different theoretical results and the NIST values
		are listed along with the present theoretical lifetimes ($\tau$, in s). The NIST compiled values reported in square brackets are determined by semi-empirical parametric calculations using Cowan's code. The rest of the NIST values are deduced from measured lines.
		The number reported in parenthesis, after the NIST excitation energy, is the estimated accuracy provided by the NIST ASD. The results from the previous calculations~\cite{Aggarwal.2016.V111-112.p187,Xu.2017.V95.p283} that do not correspond to
			this order are highlighted in boldface.
		\label{Table3}}\\
	\hline\hline\\
	\multirow{2}{*}{Key} & \multirow{2}{*}{Level} &\multicolumn{5}{c}{$\te{$E$\ (cm$^{-1}$)}$} 
	&\multicolumn{5}{c}{$\te{$\Delta E=E_{\rm{theory}}-E_{\rm{NIST}}$\ (cm$^{-1}$)}$} & \multirow{2}{*}{$\tau$ (s)}\\
	\cline{3-7} \cline{9-12} \\
	& & \tiny NIST & \tiny This work &  \tiny Aggarwal1 & \tiny  Aggarwal2  & \tiny Xu &  & \tiny This work  & \tiny  Aggarwal1 & \tiny  Aggarwal2     & \tiny Xu  &    \\
	\hline \\
	\endfirsthead\\
	\caption[]{(\emph{continued})}\\
	\hline\hline\\
	
	\multirow{2}{*}{Key} & \multirow{2}{*}{Level} &\multicolumn{5}{c}{$\te{$E$\ (cm$^{-1}$)}$} &\multicolumn{5}{c}{$\te{$\Delta E$\ (cm$^{-1}$)}$} & \multirow{2}{*}{$\tau$ (s)}\\
	\cline{3-7} \cline{9-12} \\
	& & \tiny NIST & \tiny This work &  \tiny Aggarwal1 & \tiny  Aggarwal2  & \tiny Xu &  & \tiny This work  & \tiny  Aggarwal1 & \tiny  Aggarwal2     & \tiny Xu  &    \\
	\hline\\
	\endhead
	1	&	$	     3s^23p^4$ $^3P_{2}      		$	&	0	&	0	&	0	&	0	&	0	&		&	0	&	0	&	0	&	0	&	$		$	\\	
	2	&	$	 3s^23p^4$ $^1S_{0}      		$	&	[153000](10000)	&	159769	&	161215	&	161040	&	155262	&		&	6769	&	8215	&	8040	&	2262	&	$	6.98 \rm{E}-02	$	\\	
	3	&	$	   3s^23p^3(^2P)3d\ ^3D_{2}    	^o	$	&	1893900(700)	&	1893744	&	1897578	&	1889677	&	1881974	&		&	-156	&	3678	&	-4223	&	-11926	&	$	1.14 \rm{E}-10	$	\\	
	4	&	$	 3s^23p^3(^2P)3d\ ^3P_{1}  	^o	$	&	[1959000](20000)	&	1962235	&	1966635	&	1958690	&		&		&	3235	&	7635	&	-310	&		&	$	5.08\rm{E}-11	$	\\	
	5	&	$	 3s^23p^3(^2P)3d\ ^3P_{0}    	^o	$	&	[1959000](20000)	&	1964560	&	1968830	&	1960885	&	\textbf{1954201}	&		&	5560	&	9830	&	1885	&	-4799	&	$	1.48\rm{E}-01	$	\\	
	6	&	$	  3s^23p^3(^2P)3d\ ^3F_{3}  	^o	$	&	1966200(800)	&	1966449	&	1971288	&	1963343	&	\textbf{1953805}	&		&	249	&	5088	&	-2857	&	-12395	&	$	6.19\rm{E}-10	$	\\	
	7	&	$	    3s^23p^3(^2P)3d\ ^3F_{4}    	^o	$	&	2574320(110)	&	2574205	&	2576171	&	2568896	&	2559173	&		&	-115	&	1851	&	-5424	&	-15147	&	$	3.81\rm{E}-07	$	\\	
	8	&	$	    3s^23p^3(^2P)3d\ ^3P_{2}    	^o	$	&	[2627000](30000)	&	2627280	&	2629712	&	2622448	&	2640103	&		&	280	&	2712	&	-4552	&	13103	&	$	6.99\rm{E}-12	$	\\	
	9	&	$	   3s^23p^3(^2P)3d\ ^3D_{3}  	^o	$	&	[2775000](30000)	&	2780684	&	2786483	&	2779251	&	2781879	&		&	5684	&	11483	&	4251	&	6879	&	$	1.96\rm{E}-12	$	\\	
	10	&	$	  3s^23p^4$ $^3P_{1}    		$	&		&	2798651	&	2797193	&	2799772	&		&		&		&		&		&		&	$	2.17 \rm{E}-09	$	\\	
	11	&	$	 3s^23p^3(^2P)3d\ ^1P_{1}  	^o	$	&	[2849000](30000)	&	2871998	&	\textbf{2878201}	&	2870904	&	2873826	&		&	22998	&	29201	&	21904	&	24826	&	$	3.10 \rm{E}-12	$	\\	
	12	&	$	    3s^23p^4$ $^1D_{2}      		$	&		&	2875791	&	\textbf{2875524}	&	2877949	&		&		&		&		&		&		&	$	3.41 \rm{E}-09	$	\\	
	13	&	$	 3s^23p^2(^3P)3d^2(^3F_2)\ ^5G_{2}  		$	&		&	3867430	&	3877469	&	3860756	&		&		&		&		&		&		&	$	3.53 \rm{E}-11	$	\\	
	14	&	$	3s^23p^2(^3P)3d^2(^3P_2)\ ^5D_{0}  		$	&		&	4063197	&	4074064	&	4057197	&		&		&		&		&		&		&	$	2.50 \rm{E}-11	$	\\	
	15	&	$	   3s3p^5\ ^3P_{2}    	^o	$	&	4282700(1800)	&	4282365	&	4295404	&	4284562	&	4276537	&		&	-335	&	12704	&	1862	&	-6163	&	$	1.16 \rm{E}-12	$	\\	
	16	&	$	   3s3p^5\ ^1P_{1}  	^o	$	&	[4458000](40000)	&	4450056	&	4460438	&	4451395	&	4500816	&		&	-7944	&	2438	&	-6605	&		&	$	9.84 \rm{E}-13	$	\\	
	17	&	$	3s^23p^2(^3P)3d^2(^3F_2)\ ^5G_{3}		$	&		&	4580627	&	4589730	&	4573577	&		&		&		&		&		&		&	$	3.99 \rm{E}-12	$	\\	
	18	&	$	  3s^2 3p^3(^2D)3d\ ^3F_{2}    	^o	$	&	[4656000](50000)	&	4614513	&	4615925	&	4610613	&		&		&	-41487	&	-40075	&	-45387	&		&	$	1.09 \rm{E}-10	$	\\	
	19	&	$	 3s^2 3p^3(^4S)3d\ ^5D_{0}    	^o	$	&	[4667000](50000)	&	4626085	&	4627392	&	4622092	&		&		&	-40915	&	-39608	&	-44908	&		&	$	7.14 \rm{E}-11	$	\\	
	20	&	$	  3s^2 3p^3(^2D)3d\ ^3D_{1}  	^o	$	&	[4675000](50000)	&	4635782	&	4640056	&	4633450	&		&		&	-39218	&	-34944	&	-41550	&		&	$	2.82 \rm{E}-12	$	\\	
	21	&	$	 3s^23p^2(^1S)3d^2(^3P_2)\ ^3P_{2}  		$	&		&	4670360	&	4680582	&	4664407	&		&		&		&		&		&		&	$	3.50 \rm{E}-12	$	\\	
	22	&	$	3s^2 3p^3(^2D)3d\ ^3G_{3}  	^o	$	&	[4721000](50000)	&	4679219	&	4681745	&	4676379	&		&		&	-41781	&	-39255	&	-44621	&		&	$	2.11 \rm{E}-11	$	\\	
	23	&	$	3s^23p^2(^3P)3d^2(^3P_2)\ ^5D_{1}		$	&		&	4708593	&	4718968	&	4702760	&		&		&		&		&		&		&	$	3.30 \rm{E}-12	$	\\	
	24	&	$	3s^23p^2(^1S)3d^2(^1G_2)\ ^1G_{4}  		$	&		&	4711168	&	4723017	&	4706875	&		&		&		&		&		&		&	$	3.10 \rm{E}-12	$	\\	
	25	&	$	3s^2 3p^3(^2D)3d\ ^3G_{4}    	^o	$	&	[4790000](50000)	&	4745009	&	4749080	&	4743582	&		&		&	-44991	&	-40920	&	-46418	&		&	$	1.47 \rm{E}-08	$	\\	
	26	&	$	  3s^23p^3(^2P)3d\ ^3F_{2}    	^o	$	&	[4891000](50000)	&	4852399	&	4857270	&	4851640	&		&		&	-38601	&	-33730	&	-39360	&		&	$	3.82 \rm{E}-12	$	\\	
	27	&	$	3s^2 3p^3(^2D)3d\ ^1F_{3}  	^o	$	&	4963500(1500)	&	4964223	&	4973043	&	4967984	&	4973658	&		&	723	&	9543	&	4484	&		&	$	3.46 \rm{E}-13	$	\\	
	28	&	$	3s^2 3p^3(^2D)3d\ ^3P_{2}    	^o	$	&	5018300(2500)	&	5020965	&	5032784	&	5026803	&	4963842	&		&	2665	&	14484	&	8503	&		&	$	2.44 \rm{E}-13	$	\\	
	29	&	$	3s^2 3p^3(^2D)3d\ ^3S_{1}  	^o	$	&	5062800(2100)	&	5065599	&	5078500	&	5072377	&		&		&	2799	&	15700	&	9577	&		&	$	2.11 \rm{E}-13	$	\\	
	30	&	$	3s^23p^3(^2P)3d\ ^3D_{1}  	^o	$	&	[5170000](50000)	&	5169952	&	5184737	&	5178032	&	5144567	&		&	-48	&	14737	&	8032	&		&	$	2.01 \rm{E}-13	$	\\	
	31	&	$	3s^2 3p^3(^2D)3d\ ^3D_{3}  	^o	$	&	[5299000](50000)	&	5261355	&	5260412	&	5255781	&		&		&	-37645	&	-38588	&	-43219	&		&	$	6.92 \rm{E}-11	$	\\	
	32	&	$	3s^2 3p^3(^4S)3d\ ^5D_{4}    	^o	$	&	[5299000](50000)	&	5263768	&	5263078	&	5258459	&		&		&	-35232	&	-35922	&	-40541	&		&	$	2.79 \rm{E}-09	$	\\	
	33	&	$	3s^2 3p^3(^2D)3d\ ^3F_{4}    	^o	$	&	[5406000](50000)	&	5364720	&	5365562	&	5360811	&		&		&	-41280	&	-40438	&	-45189	&		&	$	9.71 \rm{E}-09	$	\\	
	34	&	$	3s^23p^2(^1S)3d^2(^3F_2)\ ^3F_{4}  		$	&		&	5367176	&	5376777	&	5361206	&		&		&		&		&		&		&	$	2.12 \rm{E}-12	$	\\	
	35	&	$	3s^2 3p^3(^2D)3d\ ^3G_{5}  	^o	$	&	[5428000](50000)	&	5389990	&	\textbf{5391592}	&	\textbf{5386840}	&		&		&	-38010	&	-36408	&	-41160	&		&	$	1.11 \rm{E}-08	$	\\	
	36	&	$	3s^2 3p^3(^4S)3d\ ^3D_{1}  	^o	$	&	[5420000](50000)	&	5390009	&	\textbf{5390857}	&	\textbf{5386149}	&		&		&	-29991	&	-29143	&	-33851	&		&	$	8.64 \rm{E}-12	$	\\	
	37	&	$	3s^2 3p^3(^2D)3d\ ^1S_{0}    	^o	$	&	[5447000](50000)	&	5408130	&	5408919	&	5404102	&		&		&	-38870	&	-38081	&	-42898	&		&	$	1.15 \rm{E}-11	$	\\	
	38	&	$	3s^23p^2(^3P)3d^2(^1D_2)\ ^3F_{2}  		$	&		&	5442097	&	5452375	&	5436804	&		&		&		&		&		&		&	$	1.76 \rm{E}-12	$	\\	
	39	&	$	3s^2 3p^3(^4S)3d\ ^3D_{2}    	^o	$	&	[5562000](60000)	&	5538838	&	5541306	&	5536467	&		&		&	-23162	&	-20694	&	-25533	&		&	$	4.25 \rm{E}-11	$	\\	
	40	&	$	3s^23p^3(^2P)3d\ ^1F_{3}  	^o	$	&	[5620000](60000)	&	5594869	&	5598458	&	5593585	&		&		&	-25131	&	-21542	&	-26415	&		&	$	4.99 \rm{E}-10	$	\\	
	41	&	$	3s^2 3p^3(^2D)3d\ ^3D_{2}    	^o	$	&	[5643000](60000)	&	5620665	&	5626024	&	5621360	&		&		&	-22335	&	-16976	&	-21640	&		&	$	1.83 \rm{E}-12	$	\\	
	42	&	$	3s^2 3p^3(^2D)3d\ ^1P_{1}  	^o	$	&	[5674000](60000)	&	5649559	&	5656607	&	\textbf{5651505}	&		&		&	-24441	&	-17393	&	-22495	&		&	$	1.28 \rm{E}-12	$	\\	
	43	&	$	3s^23p^2(^3P)3d^2(^1S_0)\ ^3P_{0}  		$	&		&	5653246	&	5665277	&	\textbf{5649639}	&		&		&		&		&		&		&	$	1.62 \rm{E}-12	$	\\	
	44	&	$	3s^2 3p^3(^4S)3d\ ^5D_{3}  	^o	$	&	[5718000](60000)	&	5688356	&	5694906	&	5690165	&		&		&	-29645	&	-23094	&	-27835	&		&	$	1.50 \rm{E}-12	$	\\	
	45	&	$	3s^2 3p^3(^2D)3d\ ^1D_{2}    	^o	$	&	[5751000](60000)	&	5720071	&	5728804	&	5723613	&		&		&	-30929	&	-22196	&	-27387	&		&	$	1.04 \rm{E}-12	$	\\	
	46	&	$	      3s^23p^4$ $^3P_{0}      		$	&		&	5733781	&	5732414	&	5737308	&		&		&		&		&		&		&	$	3.23 \rm{E}-10	$	\\	
	47	&	$	 3s3p^4(4P)3d\ ^5D_{2}  		$	&		&	6039589	&	6055667	&	6037133	&		&		&		&		&		&		&	$	3.10 \rm{E}-12	$	\\	
	48	&	$	 3s3p^4(4P)3d\ ^5P_{1}		$	&		&	6075531	&	6092341	&	6073796	&		&		&		&		&		&		&	$	2.21 \rm{E}-12	$	\\	
	49	&	$	 3s3p^4(4P)3d\ ^5D_{3}		$	&		&	6102463	&	6119776	&	6101055	&		&		&		&		&		&		&	$	2.40 \rm{E}-12	$	\\	
	50	&	$	 3s3p^4(4P)3d\ ^3F_{4}  		$	&		&	6166554	&	6186046	&	6167160	&		&		&		&		&		&		&	$	1.77 \rm{E}-12	$	\\	
	51	&	$	 3s3p^4(2S)3d\ ^3D_{1}		$	&		&	6290183	&	6308568	&	6290066	&		&		&		&		&		&		&	$	9.81 \rm{E}-13	$	\\	
	52	&	$	 3s3p^4(2P)3d\ ^3P_{0}  		$	&		&	6362557	&	6381510	&	6362800	&		&		&		&		&		&		&	$	8.87 \rm{E}-13	$	\\	
	53	&	$	3s3p^4(2S)3d\ ^3D_{2}  		$	&		&	6388063	&	6404204	&	6387568	&		&		&		&		&		&		&	$	1.01 \rm{E}-12	$	\\	
	54	&	$	 3s3p^4(2D)3d\ ^3G_{3}		$	&		&	6412456	&	6433657	&	6415156	&		&		&		&		&		&		&	$	5.61 \rm{E}-13	$	\\	
	55	&	$	3s3p^4(4P)3d\ ^5F_{1}		$	&		&	6417641	&	6434437	&	6417427	&		&		&		&		&		&		&	$	1.16 \rm{E}-12	$	\\	
	56	&	$	3s3p^4(2P)3d\ ^3F_{2}  		$	&		&	6429402	&	6450228	&	6431079	&		&		&		&		&		&		&	$	8.72 \rm{E}-13	$	\\	
	57	&	$	3s^23p^2(^3P)3d^2(^1D_2)\ ^3F_{3}		$	&		&	6655158	&	6667562	&	6652441	&		&		&		&		&		&		&	$	7.12 \rm{E}-13	$	\\	
	58	&	$	3s^23p^2(^3P)3d^2(^3F_2)\ ^5F_{1}		$	&		&	6662143	&	6679074	&	6662130	&		&		&		&		&		&		&	$	4.56 \rm{E}-13	$	\\	
	59	&	$	3s^23p^2(^3P)3d^2(^3F_2)\ ^5F_{2}  		$	&		&	6675988	&	6695447	&	6677900	&		&		&		&		&		&		&	$	3.34 \rm{E}-13	$	\\	
	60	&	$	3s^23p^2(^1D)3d^2(^3F_2)\ ^3H_{4}  		$	&		&	6694464	&	6706124	&	6691507	&		&		&		&		&		&		&	$	1.20 \rm{E}-12	$	\\	
	61	&	$	3s3p^4(4P)3d\ ^5F_{5}		$	&		&	6704396	&	6720127	&	6701844	&		&		&		&		&		&		&	$	2.88 \rm{E}-12	$	\\	
	62	&	$	3s3p^4(4P)3d\ ^5D_{4}  		$	&		&	6728314	&	6744203	&	6725987	&		&		&		&		&		&		&	$	2.07 \rm{E}-12	$	\\	
	63	&	$	3s3p^4(4P)3d\ ^3P_{0}  		$	&		&	6878106	&	6895202	&	6877270	&		&		&		&		&		&		&	$	1.78 \rm{E}-12	$	\\	
	64	&	$	3s3p^4(4P)3d\ ^3P_{1}		$	&		&	6914978	&	6932666	&	6914592	&		&		&		&		&		&		&	$	1.16 \rm{E}-12	$	\\	
	65	&	$	3s3p^4(4P)3d\ ^3F_{3}		$	&		&	6926242	&	6943508	&	6925621	&		&		&		&		&		&		&	$	1.42 \rm{E}-12	$	\\	
	66	&	$	3s^23p^2(^3P)3d^2(^3P_2)\ ^5P_{1}		$	&		&	6960549	&	6977175	&	6962503	&		&		&		&		&		&		&	$	2.79 \rm{E}-13	$	\\	
	67	&	$	3s^23p^2(^1D)3d^2(^3F_2)\ ^3F_{2}  		$	&		&	6981515	&	6997971	&	6983782	&		&		&		&		&		&		&	$	2.46 \rm{E}-13	$	\\	
	68	&	$	3s3p^4(2P)3d\ ^3F_{4}  		$	&		&	7030629	&	7047276	&	7029970	&		&		&		&		&		&		&	$	8.87 \rm{E}-13	$	\\	
	69	&	$	 3s3p^4(4P)3d\ ^5P_{3}		$	&		&	7043764	&	7059215	&	7045399	&		&		&		&		&		&		&	$	2.95 \rm{E}-13	$	\\	
	70	&	$	3s3p^4(2S)3d\ ^3D_{3}		$	&		&	7057111	&	\textbf{7079868}	&	7060740	&		&		&		&		&		&		&	$	2.97 \rm{E}-13	$	\\	
	71	&	$	3s^23p^2(^3P)3d^2(^3F_2)\ ^5D_{1}		$	&		&	7057700	&	\textbf{7078079}	&	7063176	&		&		&		&		&		&		&	$	1.58 \rm{E}-13	$	\\	
	72	&	$	3s3p^4(4P)3d\ ^3D_{2}  		$	&		&	7067673	&	7092290	&	7070331	&		&		&		&		&		&		&	$	1.03 \rm{E}-12	$	\\	
	73	&	$	3s^23p^2(^1D)3d^2(^1S_0)\ ^1D_{2}  		$	&		&	7080981	&	7097206	&	7086090	&		&		&		&		&		&		&	$	2.00 \rm{E}-13	$	\\	
	74	&	$	   3s3p^5\ ^3P_{0}    	^o	$	&	[7141000](70000)	&	7094887	&	7104196	&	7096811	&		&		&	-46113	&	-36804	&	-44189	&		&	$	7.70 \rm{E}-13	$	\\	
	75	&	$	  3s^23p^2(^1D)3d^2(^3F_2)\ ^3P_{0}         		$	&		&	7115762	&	7137216	&	7122182	&		&		&		&		&		&		&	$	1.36 \rm{E}-13	$	\\	
	76	&	$	3s3p^4(2P)3d\ ^3D_{3}		$	&		&	7121693	&	7141167	&	7124114	&		&		&		&		&		&		&	$	5.86 \rm{E}-13	$	\\	
	77	&	$	 3s3p^4(2S)3d\ ^1D_{2}  		$	&		&	7158298	&	7176985	&	7159614	&		&		&		&		&		&		&	$	1.04 \rm{E}-12	$	\\	
	78	&	$	3s^23p^2(^3P)3d^2(^3F_2)\ ^5G_{4}  		$	&		&	7196198	&	\textbf{7217588}	&	7188837	&		&		&		&		&		&		&	$	3.58 \rm{E}-12	$	\\	
	79	&	$	 3s3p^4(4P)3d\ ^3D_{1}		$	&		&	7197873	&	\textbf{7203212}	&	7200249	&		&		&		&		&		&		&	$	6.60 \rm{E}-13	$	\\	
	80	&	$	3s^23p^2(^1D)3d^2(^3F_2)\ ^3H_{5}		$	&		&	7227799	&	7234235	&	7220397	&		&		&		&		&		&		&	$	9.19 \rm{E}-11	$	\\	
	81	&	$	3s^23p^2(^3P)3d^2(^3P_2)\ ^5D_{3}		$	&		&	7240569	&	7248995	&	7234048	&		&		&		&		&		&		&	$	1.88 \rm{E}-12	$	\\	
	82	&	$	3s^23p^2(^3P)3d^2(^3P_2)\ ^5D_{2}  		$	&		&	7240809	&	7252912	&	7238350	&		&		&		&		&		&		&	$	1.24 \rm{E}-12	$	\\	
	83	&	$	3s^23p^2(^3P)3d^2(^3P_2)\ ^5S_{2}  		$	&		&	7275238	&	7281839	&	7268254	&		&		&		&		&		&		&	$	5.78 \rm{E}-12	$	\\	
	84	&	$	3s^23p^2(^3P)3d^2(^3P_2)\ ^3P_{1}		$	&		&	7296416	&	7303688	&	7289597	&		&		&		&		&		&		&	$	3.54 \rm{E}-12	$	\\	
	85	&	$	   3s3p^5\ ^3P_{1}  	^o	$	&	[7345000](70000)	&	7298774	&	7313827	&	7301372	&		&		&	-46226	&	-31173	&	-43628	&		&	$	3.82 \rm{E}-13	$	\\	
	86	&	$	3s^23p^2(^3P)3d^2(^3P_2)\ ^3P_{0}  		$	&		&	7307588	&	7315693	&	7305411	&		&		&		&		&		&		&	$	1.49 \rm{E}-12	$	\\	
	87	&	$	3s^23p^2(^3P)3d^2(^3F_2)\ ^3G_{3}		$	&		&	7334991	&	7343753	&	7329783	&		&		&		&		&		&		&	$	3.02 \rm{E}-12	$	\\	
	88	&	$	3s^23p^2(^1D)3d^2(^3P_2)\ ^3D_{2}  		$	&		&	7339300	&	7356888	&	7338815	&		&		&		&		&		&		&	$	8.94 \rm{E}-13	$	\\	
	
	\hline
\end{longtable*}

$^{}$ \\

\clearpage
\clearpage
\setlength{\tabcolsep}{0.5\tabcolsep}
\scriptsize

\begin{longtable*}{@{\extracolsep\fill}llccccccrrrr@{}}
	\caption{Comparison of the  present MCDHF/RCI (DCTP Hamiltonian + QED-M2) wavelengths with the measured values~\cite{Ralchenko.2008.V41.p21003,Lennartsson.2013.V87.p62505,Clementson.2010.V81.p52509}, and with previous theoretical results (Aggarwal1 and Aggarwal2~\cite{Aggarwal.2016.V111-112.p187}, and Xu~\cite{Xu.2017.V95.p283}). The deviations $\Delta \lambda$ (in m{\AA}) of the different theoretical values from the experimental wavelengths are also listed.
		\label{Table4}}\\
	\hline\hline\\
	\multirow{1}{*}{Upper level} &  \multirow{1}{*}{Lower level} & \multirow{1}{*}{TM} & \multicolumn{5}{c}{$\te{$\lambda$}$~(in \AA)} &\multicolumn{4}{c}{$\te{$\Delta \lambda$}$~(in m{\AA})} \\
	\cline{4-8}  \cline{9-12} \\
	& & & Exp.  &  This work & Aggarwal1 & Aggarwal2& Xu &  This work & Aggarwal1 & Aggarwal2& Xu\\
	\hline \\
	\endfirsthead\\
	\caption[]{(\emph{continued})}\\
	\hline\hline\\
	\multirow{1}{*}{UL} &  \multirow{1}{*}{LL} & \multirow{1}{*}{Type} & \multicolumn{5}{c}{$\te{$\lambda$}$} &\multicolumn{4}{c}{$\te{$\delta \lambda$}$} \\
	\cline{4-8}  \cline{9-12} \\
	& & & Exp.  &  MCDHF & Aggarwal1 & Aggarwal2& Xu &  MCDHF & Aggarwal1 & Aggarwal2& Xu\\
	\hline\\
	\endhead
	$	3s^23p^3(^2D)3d$ $^3S_{1}	^o	$	&	$	3s^23p^4$ $^3P_{2}	$	&	E1	&	$	19.752(8)	^a	$	&	19.741 	&	19.69	&	19.71 	&		&	-11	&	-62 	&	-37 	&		\\
	$	3s^23p^3(^2D)3d$ $^3P_{2}	^o	$	&	$	3s^23p^4$ $^3P_{2}	$	&	E1	&	$	19.927(10)	^a	$	&	19.916 	&	19.87	&	19.89 	&	20.15 	&	-11	&	-57 	&	-34 	&	219 	\\
	$	3s^23p^3(^2D)3d$ $^1F_{3}	^o	$	&	$	3s^23p^4$ $^3P_{2}	$	&	E1	&	$	20.147(6)	^a	$	&	20.144 	&	20.11	&	20.13 	&	20.11 	&	-3	&	-37 	&	-18 	&	-41 	\\
	$	3s3p^5$ $^3P_{2}	^o	$	&	$	3s^23p^4$ $^3P_{2}	$	&	E1	&	$	23.350(1)	^a	$	&	23.352 	&	23.28	&	23.34 	&	23.38 	&	2	&	-70 	&	-10 	&	33 	\\
	$	3s^23p^4$ $^1D_{2}		$	&	$	3s^23p^4$ $^3P_{2}	$	&	M1	&	$	34.779(4)	^b	$	&	34.773 	&	34.78	&	34.75 	&		&	-6	&	1 	&	-32 	&		\\
	$	3s^23p^3(^2P)3d$ $^3D_{3}	^o	$	&	$	3s^23p^4$ $^3P_{2}	$	&	E1	&	$	35.974(2)	^b	$	&	35.962 	&	35.89	&	35.98 	&	35.95 	&	-12	&	-84 	&	7 	&	-27 	\\
	$	3s^23p^3(^2P)3d$ $^1P_{1}	^o	$	&	$	3s^23p^4$ $^1S_{0}	$	&	E1	&	$	36.881(3)	^b	$	&	36.870 	&	36.81	&	36.90 	&	36.78 	&	-11	&	-71 	&	21 	&	-97 	\\
	$	3s^23p^3(^2P)3d$ $^3P_{2}	^o	$	&	$	3s^23p^4$ $^3P_{2}	$	&	E1	&	$	38.072(2)	^b	$	&	38.062 	&	38.03	&	38.13 	&	37.88 	&	-10	&	-42 	&	60 	&	-195 	\\
	$	3s^23p^3(^2P)3d$ $^3F_{3}	^o	$	&	$	3s^23p^4$ $^3P_{2}	$	&	E1	&	$	50.86(2)	^c	$	&	50.853 	&	50.73	&	50.93 	&	51.18 	&	-7	&	-130 	&	74 	&	322 	\\
	$	3s^23p^3(^2P)3d$ $^3D_{2}	^o	$	&	$	3s^23p^4$ $^3P_{2}	$	&	E1	&	$	52.80(2)	^c	$	&	52.805 	&	52.70 	&	52.92 	&	53.14 	&	5	&	-100 	&	119 	&	336 	\\

	\hline
\end{longtable*}
\noindent
$^a$ From Ralchenko et al.~\cite{Ralchenko.2008.V41.p21003},\\
$^b$ From Lennartsson et al.~\cite{Lennartsson.2013.V87.p62505},\\
$^c$ From Clementson et al.~\cite{Clementson.2010.V81.p52509}.\\

\clearpage
\clearpage
\setlength{\tabcolsep}{0.5\tabcolsep}
\renewcommand{\arraystretch}{1.0}
\scriptsize
\setlength{\tabcolsep}{2pt}

\begin{longtable*}{@{\extracolsep\fill}llccccccccccccc@{}}
	\caption{The present  MCDHF/RCI (DCTP Hamiltonian + QED-M2) wavelengths ($\lambda$, in {\AA}), transition rates ($A$, in s$^{-1}$), weighted oscillator strengths ($gf$, dimensionless), and line strengths ($S$, in atomic units) for E1, E2, M1, and M2 transitions with  radiative branching ratios (BRs) larger than 0.1\% among the lowest 88 levels for S-like W. Only the results for the transitions among the 10 lowest levels are shown here for guidance regarding its form and content. Table~\ref{Table5} is available online in its entirety on the PRA website.
		\label{Table5}}\\
	\hline\hline\\
	j & i  & TM  & $\lambda$ & $A$ & $gf$ & $S$   & BRs \\
	\hline \\
	\endfirsthead\\
	\caption[]{(\emph{continued})}\\
	\hline\hline\\
	j & i  & TM  & $\lambda$ & $A$ & $gf$ & $S$   & BRs \\
	\hline\\
	\endhead
	
	2	&	1	&	E2	&	625.90 	&	$	1.432\rm{E}+01	$	&	$	8.409\rm{E}-10	$	&	$	1.228\rm{E}-03	$	&	$	1.00\rm{E}+00	$	\\
	3	&	1	&	E1	&	52.805 	&	$	8.784\rm{E}+09	$	&	$	1.836\rm{E}-02	$	&	$	3.192\rm{E}-03	$	&	$	1.00\rm{E}+00	$	\\
	4	&	1	&	E1	&	50.962 	&	$	1.452\rm{E}+10	$	&	$	1.696\rm{E}-02	$	&	$	2.846\rm{E}-03	$	&	$	7.40\rm{E}-01	$	\\
	4	&	2	&	E1	&	55.480 	&	$	5.169\rm{E}+09	$	&	$	7.156\rm{E}-03	$	&	$	1.307\rm{E}-03	$	&	$	2.63\rm{E}-01	$	\\
	5	&	1	&	M2	&	50.902 	&	$	6.151\rm{E}+00	$	&	$	2.389\rm{E}-12	$	&	$	1.410\rm{E}-04	$	&	$	8.99\rm{E}-01	$	\\
	5	&	3	&	E2	&	1412.1 	&	$	1.677\rm{E}-01	$	&	$	5.013\rm{E}-11	$	&	$	8.407\rm{E}-04	$	&	$	2.74\rm{E}-02	$	\\
	5	&	4	&	M1	&	43013 	&	$	4.395\rm{E}-01	$	&	$	1.219\rm{E}-07	$	&	$	1.297\rm{E}+00	$	&	$	7.69\rm{E}-02	$	\\
	6	&	1	&	E1	&	50.853 	&	$	1.616\rm{E}+09	$	&	$	4.387\rm{E}-03	$	&	$	7.344\rm{E}-04	$	&	$	9.98\rm{E}-01	$	\\
	7	&	1	&	M2	&	38.847 	&	$	2.054\rm{E}+04	$	&	$	4.182\rm{E}-08	$	&	$	1.097\rm{E}+00	$	&	$	7.79\rm{E}-03	$	\\
	7	&	6	&	M1	&	164.54 	&	$	2.603\rm{E}+06	$	&	$	9.508\rm{E}-05	$	&	$	3.869\rm{E}+00	$	&	$	9.93\rm{E}-01	$	\\
	8	&	1	&	E1	&	38.062 	&	$	1.431\rm{E}+11	$	&	$	1.555\rm{E}-01	$	&	$	1.948\rm{E}-02	$	&	$	1.00\rm{E}+00	$	\\
	9	&	1	&	E1	&	35.962 	&	$	5.112\rm{E}+11	$	&	$	6.938\rm{E}-01	$	&	$	8.215\rm{E}-02	$	&	$	1.00\rm{E}+00	$	\\
	10	&	1	&	M1	&	35.732 	&	$	3.268\rm{E}+08	$	&	$	1.877\rm{E}-04	$	&	$	1.658\rm{E}+00	$	&	$	7.07\rm{E}-01	$	\\
	10	&	1	&	E2	&	35.732 	&	$	5.378\rm{E}+06	$	&	$	3.088\rm{E}-06	$	&	$	8.391\rm{E}-04	$	&	$	1.19\rm{E}-02	$	\\
	10	&	2	&	M1	&	37.895 	&	$	1.143\rm{E}+08	$	&	$	7.382\rm{E}-05	$	&	$	6.917\rm{E}-01	$	&	$	2.47\rm{E}-01	$	\\
	10	&	3	&	E1	&	110.51 	&	$	1.132\rm{E}+06	$	&	$	6.220\rm{E}-06	$	&	$	2.263\rm{E}-06	$	&	$	2.17\rm{E}-03	$	\\
	10	&	4	&	E1	&	119.56 	&	$	7.495\rm{E}+06	$	&	$	4.818\rm{E}-05	$	&	$	1.896\rm{E}-05	$	&	$	1.67\rm{E}-02	$	\\
	10	&	5	&	E1	&	119.89 	&	$	6.156\rm{E}+06	$	&	$	3.979\rm{E}-05	$	&	$	1.571\rm{E}-05	$	&	$	1.29\rm{E}-02	$	\\
	
	\hline
\end{longtable*}


\end{document}